\documentclass[10pt,journal,twocolumn]{IEEEtran}
 \usepackage{latexsym}
 \usepackage{amsmath,amssymb,graphicx}
\usepackage{epsfig,times}
\usepackage{graphics,color,pstricks,pst-grad}
\usepackage{citesort}

\newtheorem{lemma}{Lemma}[section]

\newtheorem{prop}{Proposition}[section]

\newcommand{\IR}{\mathbb{R}}
\newcommand{\IZ}{\mathbb{Z}}

\newcommand{\bq}{\begin{equation}}
\newcommand{\eq}{\end{equation}}
\newcommand{\bqa}{\begin{eqnarray}}
\newcommand{\eqa}{\end{eqnarray}}
\newcommand{\ben}{\begin{enumerate}}
\newcommand{\een}{\end{enumerate}}

\newcommand{\post}[2]{
\centering
\includegraphics[width=#2cm]{#1.eps}
} 

\begin{document}

\title{Paging and Registration in Cellular Networks: Jointly
Optimal Policies and an Iterative Algorithm\\
{\small February 18, 2007}  }

\author{Bruce Hajek, Kevin Mitzel, and Sichao Yang \\
Department of Electrical and\\Computer Engineering\\
               University of Illinois at Urbana Champaign \\
               1308 W. Main Street, Urbana, IL 61801  \\
 }
 \maketitle

\begin{abstract}
This paper explores optimization of paging and registration policies
in cellular networks. Motion is modeled as a discrete-time Markov
process, and minimization of the discounted, infinite-horizon average
cost is addressed. The structure of jointly optimal paging and
registration policies is investigated through the use of dynamic
programming for partially observed Markov processes. It is shown that
there exist policies with a certain simple form that are
jointly optimal, though the dynamic programming approach does
not directly provide an efficient method to find the policies.

An iterative algorithm for policies with the simple form is
proposed and investigated. The algorithm alternates between
paging policy optimization and registration policy optimization.
It finds a pair of individually optimal policies, but an example
is given showing that the policies need not be jointly optimal.
Majorization theory and Riesz's rearrangement inequality are used
to show that jointly optimal paging and registration policies are
given for symmetric or Gaussian random walk models by the
nearest-location-first paging policy and distance threshold
registration policies.

\end{abstract}


\newcommand{\report}{report}

\section{Introduction}

The growing demand for personal communication
services is increasing the need for efficient utilization of the
limited resources available for wireless communication. In order
to deliver service to a mobile station (MS), the cellular network
must be able to track the MS as it roams. In this paper, the
problem of minimizing the cost of tracking is discussed. Two basic
operations involved in tracking the MS are {\it paging} and
{\it registration}.

There is a tradeoff
between the paging and registration costs. If the MS registers its
location within the cellular network more often, the paging costs
are reduced, but the registration costs are higher.
The traditional approach to paging and registration in cellular
systems uses {\it registration areas} which are groups of
cells.   An MS registers if and only if it
changes registration area. Thus, when there is an incoming call
directed to the MS, all the cells within its current registration
area are paged. Another method uses {\it reporting
centers}~\cite{Bar93}. An MS registers only when it enters the
cells of reporting centers, while every search for the MS is
restricted to the vicinity of the reporting center to which it
last reported.

Some dynamic registration schemes are examined in~\cite{Bar94} :
{\it time-based, movement-based}, and {\it distance-based}. 
These policies are threshold policies and the
thresholds depend on the MS motion activities.
In~\cite{MadhavapeddyBasuRoberts95}, dynamic programming is used
to determine an optimal {\it state-based} registration policy.
Work in \cite{AnjumTassiulasShayman02} considers congestion among
paging requests for multiple MSs, and considers overlapping registration
regions.  

Basic paging policies can be classified as follows:
\begin{itemize}
  \item {\it Serial Paging}. The cellular network pages the MS sequentially,
          one cell at a time.
  \item {\it Parallel Paging}. The cellular network pages the MS in
         a collection of cells simultaneously.
\end{itemize}
Serial paging policies have lower paging costs than parallel
paging policies, but at the expense of larger delay.  The method
of parallel paging is to partition the cells in a service region
into a series of indexed groups referred to as {\it paging areas}.
When a call arrives for the MS, the cells in the first
paging area are paged simultaneously in the first round and then,
if the MS is not found in the first round of paging, all the cells
in the second paging area are paged, and so on. 
Given disjoint paging areas, searching them in the order of decreasing
probabilities minimizes the the expected number of searches~\cite{Rose95}. 
This paging order is denoted as the {\it maximum-likelihood serial
paging order}. An interesting topic of paging is to design the optimal
paging areas within delay constraints~\cite{Rose95,Hass,Aky01}.
However, in this paper, we consider only serial paging polices.
\begin{figure}
    \centering
    \post{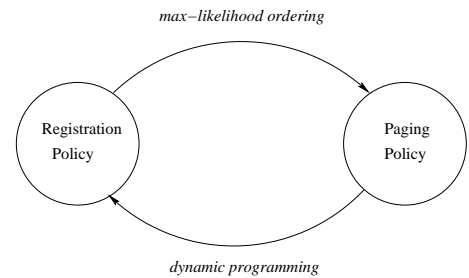}{6}
    \caption{Paging policy and registration policy generation}
    \label{fig:pair}
\end{figure}

Each paper mentioned above assumes a certain class of
paging or registration policy. Given one policy (paging
policy or registration policy) and the parameters of an assumed
motion model, the counterpart policy (registration policy or paging policy,
respectively) is found. For instance, the
optimal paging policy is identified in~\cite{Rose95} for a given registration policy. This is shown as the top branch of
Figure~\ref{fig:pair}. Conversely, an expanding ``ping-pong" order paging policy suited to the given motion model is assumed
in~\cite{MadhavapeddyBasuRoberts95}. With this knowledge, dynamic programming is applied to solve for the optimal registration policy.
This corresponds to the bottom branch of Figure~\ref{fig:pair}.

Several studies have addressed minimizing the costs, considering
the paging and registration policies
together~\cite{Rose96,Aky96,Rose99}. In~\cite{Rose96}, a
timer-based registration policy combined with maximum-likelihood
serial paging is introduced. The minimum paging cost can be
represented by the distribution of locations where the MS last
reported. Then an optimal timer threshold is selected to minimize
the total cost of registration and paging. By contrast, a
movement-based registration policy is used in~\cite{Aky96}. An
improvement of~\cite{Rose96} is given in~\cite{Rose99} by assuming
that the MS knows not only the current time, but also its own
state and the conditional distribution of its state given
the last report. This is a state-based registration policy and
is aimed to minimize the total costs by running a greedy algorithm on
the potential costs. Although the papers discuss the paging and registration
policy together, they don't consider jointly optimizing the policies.

The contributions of this paper are as follows.\footnote{Earlier versions
of this work appeared in  \cite{HajekMitzelYang03,Hajek02}. } 
The structure of jointly optimal paging and registration policies
is identified.   It is shown that the conditional probability
distribution of the states of an MS can be viewed as a controlled
Markov process, controlled by both the paging and registration
polices at each time. Dynamic programming is applied to show
that the jointly optimal policies can be represented compactly by
certain {\em reduced complexity laws} (RCLs).   An iterative algorithm
producing a pair of RCLs is proposed based on closing the loop in
Figure~\ref{fig:pair}. The algorithm is a heuristic which merges
the approaches in~\cite{MadhavapeddyBasuRoberts95}
and~\cite{Rose95}.  Several examples are given.  The first example
is an  illustration of numerical computation of an individually optimal
policy pair.   The second example is a simple one illustrating that individually
optimal policies are not necessarily jointly optimal.   Finally, three
more examples are given based on random walk models of motion:
one-dimensional discrete state symmetric, multidimensional symmetric,
and multidimensional Gaussian.     Majorization theory and
Riesz's rearrangement inequality are used to show that jointly
optimal paging and registration policies are given for these random
walk models by the nearest-location-first paging policy and distance
threshold registration policies.

The paper is organized as follows. Notation and cost
functions are introduced in Section~\ref{sec:model}.
Jointly optimal policies are investigated in Section~\ref{sec:joint}. The
iterative optimization formula for computing individually optimal
policy pairs is developed in Section~\ref{sec:individ}.
The first two examples are given in Section ~\ref{sec:examples}, and the
random walk examples are given in Section ~\ref{sec:random_walk}.
Conclusions are given in Section \ref{sec:conclusion}.

\section{NETWORK MODEL}

\label{sec:model}

\subsection{State description and cost}

Let $\cal C$ denote the set of cells, which is assumed to be finite.   A cell $c$
is a physical location that the MS can physically be in.    The motion of an MS is
modeled by a discrete-time Markov process $(X(t):t\geq 0)$ with a finite state
space $S$, one-step transition probability matrix $P=(p_{ij} : i,j \in S)$,
and given initial state $x_0$. A state $j\in S$  determines the cell $c$ that the MS is
physically located in, and it may indicate additional information, such as the current
velocity of the MS,  or the previously visited cell.   Thus a cell
$c$ can be considered to be a set of one or more states, and the
set $\cal C$ of all cells is a partition of $S$.  It is assumed
that the network knows the initial state $x_0$.   In the special case that there
is one state per cell, we write ${\cal C}=S$, and then the MS moves among the
cells according to a Markov process.

The possible events at a particular integer time instant $t\geq 1$ are as follows, listed in the
order that they can occur.   First, the state $X(t)$ is generated based on $X(t-1)$ and the
one-step transition probability matrix $P$.   Then, it is determined whether the MS is to be paged,
and the answer is ``yes'' with probability $\lambda_p$,
independently of the state of the MS and all past events.  The
cost of the paging at time $t$ is ${\cal P} N_t$, where $\cal P$
is the cost of searching one cell and $N_t$ is the number of cells
that are searched until the MS is found. If the MS is paged, the cellular network learns
the state, $X(t)$.
Let $N_t=0$ if the MS is not paged at time $t$.     Finally, if the MS was not paged, the
MS decides whether to register.  The cost of registration is $\cal R$ and the benefit of
registration is that the cellular network learns the state of the
MS.   No paging or registration is considered for $t=0$. 

Let $P_t$ denote the event that the MS is paged at time $t$, and
let $R_t$ denote the event that the MS registers at time $t$. 
We say that a {\em report} occurs at time $t$ if either a paging
or a registration occurs, because in either case, the cellular
network learns the state of the MS.  For  any set $A$, let $I_A$ denote the
indicator function of $A$, which is one on $A$ and zero on the complement $A^c$.
Discrete probability distributions are considered to be row vectors. Given a state $l\in S$,
let $\delta(l)$ denote the probability distribution on $S$ which
assigns probability one to state $l$.  Thus,
$\delta_i(l)=I_{\{i=l\}}$. See the appendix for a review of the
notions of $\sigma$-algebras used in this paper.

\subsection{Paging policy notation}

For simplicity we consider only serial paging policies, so that
cells are searched one at a time until the MS is located.
It is also assumed that if the MS is present in the cell in which it
is paged, it responds to the page successfully.  In other words,
no paging failure is allowed.  It is further assumed that the time
it takes to issue a single-cell page is negligible compared to one
time step of the MS's motion model, so that paging is always
successfully completed within one time step.

Let ${\cal N}_t$ denote the $\sigma$-algebra representing the
information available to the network by time $t$ after the paging
and registration decisions have been made and carried out. Thus,
for $t\geq 0$,
\begin{eqnarray*}
{\cal N}_t & = & \sigma ( (I_{P_s},N_s,I_{R_s}:1\leq s \leq t), \\
&& (X(s):1\leq s \leq t ~\mbox{and}~ I_{P_s \cup R_s}=1))
\end{eqnarray*}
The initial state $x_0$ is treated as a constant, so even though
it is known to the network it is not included in the definition of
${\cal N}_t$.  Note that the initial $\sigma$-algebra ${\cal N}_0$
is the trivial $\sigma$-algebra: ${\cal N}_0=\{\emptyset, \Omega \}.$

When the MS is to be paged, the cells are to be searched
sequentially according to a permutation $a$ of the cells.
The associated {\em paging order vector} $r=(r_j : j \in S)$ is such
that for each state $j$, $r_j$ is the number of cells that must be
paged until the cell for state $j$ comes up, and the MS is reached.
For example, suppose $ {S} = \{1,2,3,4,5,6\} $ and
${\cal C}=\{c_1,c_2,c_3\}$ with $c_1=\{1,2\}$, $c_2=\{3,4\}$, and
$c_3=\{5,6\}$.   Then if the cells are search according to the permutation
$a=(c_2,c_1,c_3)$, meaning to search cell
$c_2$ first, $c_1$ second, and $c_3$ third, then the paging order vector
is  $r=(2,2,1,1,3,3).$   A {\em paging policy} $u$ is a collection
$u=(u(t):t\geq 1)$ such that for each $t\geq 1$,  $u(t)$ is an
${\cal N}_{t-1}$ measurable random variable with values in the set of paging
order vectors.  Note that $N_t = I_{P_t} u_{X(t)}(t)$.

\subsection{Registration policy notation}

Let ${\cal M}_t$ denote the $\sigma$-algebra representing the
information available to the MS by time $t$, after the paging and
registration decisions for time $t$ have been made and carried
out. Thus,
\begin{displaymath}
{\cal M}_t = \sigma ( X(s), I_{P_s}, N_s, I_{R_s}:1\leq s \leq t).
\end{displaymath}
The MS also knows the initial position $x_0$, which is treated as
a constant. In practice an MS wouldn't learn $N_s$, the number of
pages used to find the MS at time $s$.  While we assume such
information is available to the MS, we will see that optimal
policies need not make use of the information. With this
definition, we have ${\cal N}_t \subset {\cal M}_t$, meaning that
the MS knows everything the network knows (and typically more).

When the MS has to decide whether to register at time $t$, it
already has the information ${\cal M}_{t-1}$.  In addition it
knows $X(t)$ and $I_{P_t}$. If the MS is paged at time $t$,
then the network learns the state of the MS as a result, so
there is no advantage for the MS to register at time $t$.  Thus,
we assume without loss of generality that the MS does not
register at time $t$ if it is paged at time $t$.  This leads to
the following definition.

A {\em registration policy} $v$ is a collection $v=(v(t):t\geq 1)$ such
that for each $t\geq 1$, $v(t)$ is an ${\cal M}_{t-1}$ measurable
random vector with values in $[0,1]^S$ with the following
interpretation.  Given the information ${\cal M}_{t-1}$, if
$X(t)=l$ and if the MS is not paged at time $t$, then the MS
registers with probability $v_l(t)$.

\subsection{Cost function}

Let $\beta$ be a number
with $0<\beta<1$, called the {\em discount factor.}  An interpretation
of $\beta$ is that $1/(1-\beta)$ is the rough time
horizon of interest.  
Given a paging policy $u$ and registration policy $v$,  the expected
infinite horizon discounted cost $C(u,v)$ is defined as
\begin{equation}
C(u,v) = E \left[ \sum_{t=1}^\infty \beta^t \{ {\cal P}I_{P_t}N_t
+ {\cal R}I_{R_t} \}  \right].
\end{equation}
The pair $(u,v)$ is {\em jointly  optimal} if $C(u,v)\leq
C(u',v')$ for every paging policy $u'$ and registration
policy $v'$.

\section{JOINTLY OPTIMAL POLICIES} \label{sec:joint}

This section investigates the structure of jointly optimal
policies by using the theory of dynamic programming for Markov
control problems with partially observed states. While the
structure results do not directly yield a computationally feasible
solution, they shed light on the nature of the problem. In
particular it is found that there are jointly optimal policies
$(u,v)$ such that, for each $t$, $u(t)$ and $v(t)$ are functions of
the amount of time elapsed since the last \report\ and the last
reported state.

Intuitively, on one hand, the paging policies are selected based on the past of
the registration policy, because the past of the registration
policy influences the conditional distribution of the MS
state.  On the other hand, by the nature of dynamic programming,
the optimal choice of registration policy at a given time depends
on future costs, which are determined by the future of the
registration policy. To break this cycle, we consider the problem
entirely from the viewpoint of the network.  In order that current
decisions not depend on past actions, the state space is augmented
by the conditional distribution of the state of the MS given the
information available to the network.

\subsection{Evolution of conditional distributions}

For $t\geq 0$, let $w(t)$ be the conditional
probability distribution of $X(t)$, given the observations
available to the network up to time $t$ (including the outcomes of
a \report\ at time $t$, if there was any). That is,
$w_j(t)=P[X(t)=j|{\cal N}_t]$ for $j\in S$. Note that, with
probability one, $w(t)$ is a probability distribution on $S.$
Intuitively, the network can control the distribution
valued process $(w(t))$ by dictating the registration policy of
the MS. Since ${\cal N}_0$ is the trivial $\sigma$-algebra and
$X(0)=x_0$, the initial conditional distribution is given by
$w(0)=\delta(x_0)$.

While the network may not know the recent past trajectory of the
state process, it can still estimate the registration probabilities
used by the MS.  In particular, as shown in the next lemma,
the estimate $\widehat{v}_j(t)$, defined by
$\widehat{v}_j(t) = E[v_j(t)|  X(t)=j, {\cal N}_{t-1}],$
plays a role in how the network can recursively update the $w(t)$'s. 
In more conventional notation, we have
$$
\widehat{v}_j(t) = \frac{E[v_j(t) I_{\{X(t)=j\} } |  {\cal N}_{t-1}]  }{P[X(t)=j |  {\cal N}_{t-1}] }.
$$

Define a function $\Phi$ as follows.  Let $w$ be a probability
distribution on $S$ and let $d \in [0,1]^S$.  Let $\Phi(w,d)$
denote the probability distribution on $S$ defined by
\begin{displaymath}
\Phi_l(w,d) = \frac{ \sum_{j\in S} w_j p_{jl}(1-d_l) }{\sum_{l'\in
S}
      \sum_{j\in S} w_j p_{jl'}(1-d_{l'}) }.
\end{displaymath}
$\Phi(w,d)$ is undefined if the denominator in this definition is
zero.  The meaning of $\Phi$ is that if at time $t$ the network
knows that $X(t)$ has distribution $w$, if no paging occurs at
time $t+1$, and if the MS registers at time $t+1$ with probability
$d_{X(t+1)}$, then $\Phi(w,d)$ is the conditional distribution of
$X(t+1)$ given no registration occurs at time $t+1$.  This
interpretation is made precise in the next lemma.
The proof is in the appendix.

\begin{lemma} \label{lemma:w_rep}
The following holds, under the paging and registration policies $u$ and $v$:
\begin{eqnarray} \label{eq:w_rep}
w(t+1) &= &\delta(X(t+1))I_{P_{t+1}\cup R_{t+1}} \\
&& + \Phi(w(t),\widehat{v}(t+1))I_{P^c_{t+1}\cap R^c_{t+1}}.  \nonumber
\end{eqnarray}
\end{lemma}

\subsection{New state process}

For $t\geq 1$, let $\Theta(t) = (w(t), I_{P_t},N_t,I_{R_t})$. Note
that the $t$th term in the cost function is a function of
$\Theta(t)$.  Note also that $\Theta(t)$ is measurable with
respect to ${\cal N}_t$, so that the network can calculate
$\Theta(t)$ at time $t$ (after possible paging and registration).
Moreover, the first coordinate of $\Theta(t)$, namely $w(t)$, can
be updated with increasing $t$ with the help of Lemma
\ref{lemma:w_rep}. The random process $(\Theta(t):t\geq 0)$ can be
viewed as a controlled Markov process, adapted to the family of
$\sigma$-algebras $({\cal N}_t:t\geq 0)$ with controls
$(u(t),\widehat{v}(t): t\geq 1)$.  Note that $u(t+1)$ and
$\widehat{v}(t+1)$ are each ${\cal N}_t$ measurable for each $t\geq
0$.  The one-step transition probabilities for  $(\Theta(t))$ are
given as follows.  (The variables $j$ and $l$ range over the set
of states $S$.)

{\small
\begin{displaymath}
\begin{array}{c|c}  
\Theta(t+1)  &  \mbox{probability}  \\ \hline
(\delta(l),1,u_l (t+1) ,0) & \lambda_p \sum_j w_j(t)p_{jl} \\\\
(\delta(l),0,0,1)   &
     (1-\lambda_p) \sum_j w_j(t)p_{jl}\widehat{v}_l(t+1) \\\\
(\Phi(w(t),\widehat{v}(t+1)),0,0,0)  & 
     (1-\lambda_p) \sum_j w_j(t)p_{jl}(1-\widehat{v}_l(t+1))
\end{array}
\end{displaymath}
}

Observe that although the MS uses a registration policy $v$, the one-step transition
probabilities for $\Theta$ depend only on $\widehat{v}.$ 
Moreover,  $\widehat{v}$ is itself a
registration policy. Indeed, since ${\cal N}_{t-1} \subset {\cal
M}_{t-1}$,  $\widehat{v}(t)$ is ${\cal M}_{t-1}$ measurable, and it
takes values in $[0,1]^S$. 
If $\widehat{v}$ were used instead of $v$ as a registration
policy by the MS, the one-step transition
probabilities for $\Theta$ would be unchanged.
Thus, the policy $\widehat{v}$ is adapted to the family of $\sigma$
algebras $({\cal N}_t:t\geq 0)$, and it yields the same cost as $v$.
Therefore, without loss of generality, we can restrict attention
to registration policies $\widehat{v}$ that are adapted to $({\cal N}_t:t\geq 0)$.

Combining the observations summarized in this section, we arrive
at the following proposition.

\begin{prop}
The original joint optimization problem is equivalent to a Markov
optimal control problem with state process $(\Theta(t):t\geq 0)$
adapted to the family of $\sigma$-algebras $({\cal N}_t:t\geq 0)$,
with controls $(u(t),\widehat{v}(t): t\geq 1)$.
\end{prop}


\subsection{Dynamic programming equations}

Above it was assumed that $w(0)=\delta(x_0)$, where $x_0$ is the
initial state of the MS, assumed known by the network.   In order
to apply the dynamic programming technique, in this section the initial distribution $w(0)$ is allowed to be any probability
distribution on $S$.   It is assumed that the network knows $w(0)$
at time zero, and that the initial state of the MS is random,
with distribution $w(0)$.   The evolution of the system as
described in the previous section is well defined for an arbitrary
initial distribution $w(0)$.  Let $E_w$ denote conditional
expectation in case the initial distribution $w(0)$ is taken to be
$w$.   The initial $\sigma$-algebra ${\cal N}_0$ is still the
trivial $\sigma$-algebra,  because $w(0)$ is treated as a given
constant.

Define the cost with $n$ steps to go as
\begin{displaymath}
U_n(w)=\min_{u,\widehat{v} } E_w[\sum_{t=1}^n \beta^t \{ {\cal
P}I_{P_t}N_t+{\cal R}I_{R_t}\}]
\end{displaymath}
Next apply the backwards solution method of dynamic programming,
by separating out the $t=1$ term in the cost for $n+1$ steps to
go.  This yields
\begin{eqnarray*}
U_{n+1}(w) & = & \min_{u,\widehat{v}} \beta \left[
\lambda_p {\cal P} \sum_j
\sum_l w_jp_{jl}  u_l(1)   \right.
  \\&& +
   (1-\lambda_p){\cal R} \sum_j \sum_l w_jp_{jl}\widehat{v}_l(1)  \\
  && \left. +  E_w[
    E_w[\sum_{t=2}^{n+1} \beta^{t-1}\{ {\cal P}I_{P_t}N_t+{\cal R}I_{R_t}\}
      |{\cal N}_1] ] \right]
\end{eqnarray*}
Note that $u(1)$ and $\widehat{v}(1)$ are both measurable with respect
to the trivial $\sigma$-algebra ${\cal N}_0$.  Therefore these
controls are constants.  Henceforth we write
$d$ for the registration decision vector
$\widehat{v}(1)$. The vector $d$ ranges over the space $[0,1]^S$.

The first sum in the expression for $U_{n+1}(w)$ involves the
control policies only through the choice of the paging order vector $u(1)$.
This sum is simply the mean number of single-cell pages required
to find the MS given that the state of the MS has distribution
given by the product $wP$, where $P$ is the matrix of state
transition probabilities.   It is well known that the optimal
search order is to first search the cell with the largest
probability, then search the cell with the second largest
probability, and so on~\cite{Rose95}. Ties can be broken arbitrarily.  The first
sum in the expression for $U_{n+1}(w)$ can thus be replaced by
$s(wP)$, where $s(q)$ denotes the mean number of single cell pages
required to find the MS given that the state of the MS has
distribution $q$ and the optimal paging policy is used.  (We remark that
Massey \cite{Massey94} explored comparisons between $s(w)$, in the case of
one state per cell, and the ordinary entropy, $H(w)=\sum_i w_i\log w_i$. 
The measure $s(w)$ was called the {\em guessing entropy} in \cite{Cachin97},
and work continues to compare it to other forms of entropy \cite{DeSantisGaggiaVaccaro01}.)

The dynamic programming equation thus becomes
\begin{equation}
\begin{array}{l}
U_{n+1}(w)  =  \beta \lambda_p {\cal P} s(wP) \\
\displaystyle{~~~  \min_d \beta \left[   \sum_j \sum_l w_jp_{jl}
   \left\{ \lambda_p U_n(\delta(l)) ~~~ \right.  \right. } \\
\displaystyle{~~~~~~ \left.     + (1-\lambda_p) d_l( {\cal R}+
         U_n(\delta(l)))  \right\}    \label{eq:DP_Un}  }\\
\displaystyle{ ~~~+ \left.
  \left( \sum_j \sum_l w_j p_{jl}(1-d_l)\right) U_n(\Phi (w,d)) \right]. }
\end{array}
\end{equation}

Formally we denote this equation as $U_{n+1}=T(U_n)$. By a
standard argument for dynamic programming with discounted cost,
$T$ has the following contraction property:
\begin{equation} \label{cp}
\sup_w  | T(U)-T(U') | \leq \beta \sup_w |U-U'|
\end{equation}
for any bounded, measurable functions $U$ and $U'$, defined on the
space of all probability distributions $w$ on $S$. Consequently
\cite{Bertsekas95a,Bertsekas95b}, there exists a unique $U_*$ such
that $T(U_*)=U_*$, and $U_n\rightarrow U$ uniformly as
$n\rightarrow\infty$.   Moreover, $U_*$ is the minimum possible
cost, and a jointly optimal pair of paging and registration
policies is given by a pair $(\bar{f},\bar{g})$ of state feedback
controls, for the state process $(w(t))$. A jointly optimal
control is given by  $u(t)=\bar{f}(w(t-1))$ and
$v(t)=\bar{g}(w(t-1))$, where $\bar{f}$ and $\bar{g}$ are
determined as follows.  For any probability distribution $w$ on
$S$, $\bar{f}(w)$ is the paging order vector for
paging the cells in order of decreasing probability under
 distribution $wP$, and $\bar{g}(w)$ is a value of
$d$ that achieves the minimum in the right hand side of
(\ref{eq:DP_Un}) with $U_n$ replaced by $U_*$. Then if there is no
\report\ at time $t+1$, the conditional distribution $w(t)$ is
updated simply by:
\begin{equation} \label{eq:wupdate}
w(t+1)=\Phi(w(t),\bar{g}(w(t)))
\end{equation}
Clearly under such stationary state feedback control laws
$(\bar{f},\bar{g})$, the process $(w(t):t\geq 0)$ is a
time-homogeneous Markov process. Note that the optimal mapping
$\bar{f}$ does not depend on $\bar{g}$.

\begin{lemma}
The registration policy $\bar{g}$ can be taken to be $\{0,1\}^S$ valued
(rather than $[0,1]^S$ valued) without loss of optimality.
\end{lemma}
\begin{proof}
It is first proved that $U_n$ is concave for any given $n\geq 0$.
Suppose $w_1$ and $w_2$ are two probability distributions on $S$,
suppose $0<\eta < 1$ and suppose $w=\eta w_1+(1-\eta)w_2$. Then
$U_n(w)$ can be viewed as the cost to go given the MS has
distribution $w_1$ with probability $\eta$ and distribution $w_2$
with probability $1-\eta$, and the network does not know which
distribution is used.   The sum $\eta U_n(w_1)+(1-\eta)U_n(w_2)$
has a similar interpretation, except the network does know which
distribution is used.  Thus, the sum is less than or equal to
$U_n(w)$, so that $U_n$ is concave.  Therefore $U_*$ is also
concave.

Given a function $H$ defined on the space of all probability
distributions for $S$, let $\tilde{H}$ be an extension of $H$ defined on
the positive quadrant $R_+^S$ as follows.   For any probability
distribution $w$ and any constant $c\geq 0$, $\tilde{H}(cw)=cH(w)$. It
is easy to show that if $H$ is concave then the extension
$\tilde{H}$ is also concave.  With this notation, the dynamic
programming equation for $U_*$ can be written as:
\begin{displaymath}
\begin{array}{l}
U_*(w)  =  \beta \lambda_p {\cal P} s(wP) \\
\displaystyle{~~~  \min_d \beta \left[   \sum_j \sum_l w_jp_{jl}
   \left\{ \lambda_p U_*(\delta(l)) ~~~ \right.  \right. } \\
\displaystyle{~~~~~~ \left.     + (1-\lambda_p) d_l( {\cal R}+
         U_*(\delta(l)))  \right\}      }\\
\displaystyle{ ~~~+  \left. \tilde{U_*}(wPdiag(1-d)) \right].} \\ 
\end{array}
\end{displaymath}
where $diag(1-d)$ is the diagonal matrix with $l$th entry $1-d_l$.
The expression to be minimized over $d$ in this equation is a
concave function of $d$, and hence the minimum of the function
occurs at one of the extreme points of $[0,1]^S$, which are just
the binary vectors $\{0,1\}^S$.  The minimizing $d$ is $\bar{g}(w)$.
This completes the proof of the lemma.
\end{proof}

\subsection{Reduced complexity laws}

Given a pair of feedback controls $(\bar{f},\bar{g})$, a more
compact representation of the controls is possible.  Indeed,
suppose the controls are used, and suppose in addition that
$X(0)=x_0$, where $x_0$ is an initial state known to the network.
Given $t\geq 1$, define $k \geq 1$ and $i_0\in S$ as follows.  If
there was a \report\ before time $t$, let $t-k$ be the time of the
last \report\ before $t$. If there was no \report\ before time $t$
let $k=t$. In either case, let $i_0=X(t-k)$.   Since the network
knows $X(t-k)$ at time $t-k$ (after possible paging or
registration), we have that $w(t-k)=\delta(i_0)$.  Since there
were no state updates during the times $t-k+1,\ldots,t-1$, it
follows that $w(t-1)$ is the result of applying the update
(\ref{eq:wupdate}) $k-1$ times, beginning with $\delta(i_0)$.
Hence, $w(t-1)$ is a function of $i_0,k$.   Moreover, since
$u(t)=\bar{f}(w(t-1))$ and $v(t)=\bar{g}(w(t-1))$, it follows that
both the paging order vector $u(t)$ and the registration decision vector
$v(t)$ are determined by $i_0$ and $k$. Let $f$ and $g$ denote the
mappings such that $u(t)=f(i_0,k)$ and $v(t)=g(i_0,k)$. Note that
$f(i_0,k)$ is a paging order vector and $g(i_0,k)\in \{0,1\}^S$ for
each $i_0$, $k$. We call the mappings $f$, $g$ {\em reduced complexity
laws} (RCLs).  We have shown the following proposition.
\begin{prop}
There is no loss in optimality for the original joint paging and
registration problem to use policies based on RCLs.
\end{prop}

\begin{figure}
\post{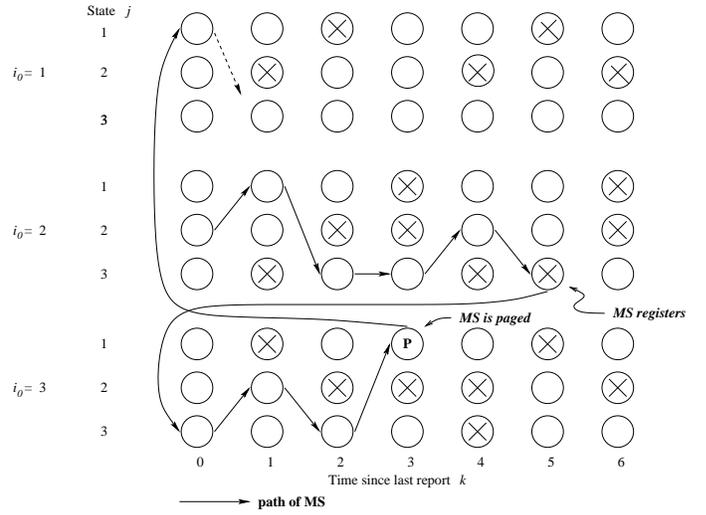}{9} \caption{Example of a registration policy represented
by an RCL for a three-state Markov chain} \label{regfig}
\end{figure}

Figure~\ref{regfig} shows an example of a registration RCL $g$ for
a three-state Markov chain.  The augmented state of a MS is a
triple $(i_0,k,j)$, such that $i_0$ is the state at the time of
the last \report, $k$ is the elapsed time since the last \report,
and $j$ is the current state. Augmented states marked with an
``$\times$'' are those for which $g_j(i_0,k)=1$, meaning that
registration occurs (if paging doesn't occur first). An MS
traverses a path from left to right until either it is paged, or
until it hits a state marked with an ``$\times$,'' at which time its
augmented state instantaneously jumps. The figure shows the path
of a MS that began in augmented state $(i_0,k,j)=(2,0,2)$.   At
relative time $k=5$ the MS entered state 3, hitting an ``$\times$'',
causing the extended state to instantly change to $(3,0,3)$. Three
time units after that, upon entering state 1, the MS is paged.
This causes the augmented state to instantly jump to $(1,0,1)$.

\section{ITERATIVE ALGORITHM FOR FINDING INDIVIDUALLY OPTIMAL POLICIES}
\label{sec:individ}

\subsection{Overview of iterative optimization formulation}

While jointly optimal policies can be efficiently represented by
RCLs $f$ and $g$,  the dynamic programming method described for
finding the optimal policies  is far from computationally
feasible, even for small state spaces, because functions of
distributions on the state space must be considered. In this
section we explore the following method for finding a pair of
policies with a certain local optimality property. First it is
show how to find, for a given paging RCL $f$, an optimal
registration RCL $g$.  Then it is shown how to find, for a given
registration RCL $g$, an optimal paging RCL $f$.   Iterating
between these two optimization problems produces a pair of RCLs
$(f,g)$ such that for each RCL fixed, the other is
optimal. Such pairs of RCLs are said to be {\em individually optimal}.

In this section we impose the constraint that an MS must
register if $k\geq k_{max}$, for some large integer constant
$k_{max}$. With this constraint, the  sets of possible
registration and paging RCLs are finite, and numerical
computation is feasible for fairly large state spaces. The initial
state $x_0$ is assumed to be known and we write $C(f,g)$ for the
averaged infinite horizon, discounted cost, for paging RCL $f$ and
registration RCL $g$.

\subsection{Optimal registration RCL for given paging RCL}
\label{reg}
Suppose a paging RCL $f$ is fixed. In this subsection we address
the problem of finding a registration RCL $g$ that minimizes
$C(f,g)$ with respect to $g$.  Dynamic programming is again used,
but here the viewpoint of the MS is taken.  The states used for
dynamic programming in this section are the augmented states of
the form $(i_0,k,j)$, rather than the set of all probability
distributions on $S$.

Since time is implicitly included in the variable $k$ in the
augmented state, it is computationally more efficient to consider
dynamic programming iterations based on cycles rather than on
single time steps, where each cycle ends when there is a \report.
Let $\tau_m$ be the time of the $m^{th}$ \report. Replacing the
infinite horizon by time horizon $\tau_m$ reduces $C(f,g)$ to
\begin{equation}
E\left[ \sum_{t=1}^{\tau_m} \beta^t \left\{ {\cal P}I_{  P_t  }N_t
 +  {\cal R}I_{R_t} \right\} \right]
\label{newcost}
\end{equation}
Letting $m \rightarrow \infty$ in (\ref{newcost}) yields $C(f,g)$.

Then for each $(i_0,j,k)$, write $V_m(i_0,k,j)$ for the {\em
cost-to-go} for $m\geq 1$ update cycles:
\begin{equation}
V_m(i_0,k,j) = \min_u E \left[ \sum_{t=1}^{\tau_m} \beta^t \left\{
{\cal P}I_{  P_t  }N_t
 +  {\cal R}I_{R_t} \right\} \right],
 \label{nctg}
\end{equation}
where the expectation $E$ is taken assuming that (a) the paging
RCL $f$ is used for the paging policy, (b) at $t=0$ the MS is in
state $j$,  and (c) the last \report\ occurred $k$ time units
earlier in state $i_0$. Also, define $V_0(i_0,k,j) \equiv 0$, because
the cost is zero when there are no \report\ cycles to go.

The dynamic programming optimality equations are given by
\begin{eqnarray}
V_m(i_0,k,j) =  \beta \sum_{l \in {S}} p_{jl} \left[
\lambda_p ({\cal P} f_l(i_0,k+1) +V_{m-1}(l,0,l)) \right. \nonumber &&\\
~~ + \left.   (1-\lambda_p)\min \left\{ V_m(i_0,k+1,l) , {\cal R} +
V_{m-1}(l,0,l) \right\}  \right] && \label{eq:dpvaluem}
\end{eqnarray}
As mentioned earlier, registration is forced at relative time
$k=k_{max}+1$ for some large but fixed value $k_{max}$.  Therefore
we set $V_m(i_0,k_{max}+1,l) =\infty$ and use (\ref{eq:dpvaluem})
only for $0\leq k \leq k_{\max}$. These equations represent the basic dynamic programming optimality
relations.  For each possible next state, the MS chooses whichever
action has lesser cost:  either continuing the current
registration cycle or registering for cost $\cal R$.

Equation (\ref{eq:dpvaluem}) can be used to compute the functions
$V_m$ sequentially in $m$ as follows. The initial conditions are
$V_0 \equiv 0$. Once $V_{m-1}$ is computed, the values
$V_m(i_0,k,j)$  can be computed using (\ref{eq:dpvaluem}),
sequentially for $k$ decreasing from $k_{\max}$ to 0. Formally we
denote this computation as $V_m=T(V_{m-1})$.  The mapping $T$ is a
contraction with constant $\beta$ in the sup norm, so that $V_m$
converges uniformly to a function $V_*$ satisfying the limiting
form of (\ref{eq:dpvaluem}):
\begin{eqnarray}
V_*(i_0,k,j)  =  \beta \sum_{l \in {S}} p_{jl} \left[
\lambda_p ({\cal P} f_l (i_0,k+1) +V_*(l,0,l)) \right.\nonumber &&\\
~~ + \left.  (1-\lambda_p)\min \left\{ V_*(i_0,k+1,l) , {\cal R} +
V_*(l,0,l) \right\}  \right] && \label{eq:dpvalue}
\end{eqnarray}
for $0\leq k \leq k_{max}$, and $V_*(i_0,k_{max}+1,l)\equiv
\infty$. The corresponding optimal registration $RCL$ $g^*$ is
given by
\begin{equation}
g^*_l(i_0,k) =  \left\{ \begin{array}{ll}
0, & \mbox{if } V_*(i_0,k+1,l) \leq  {\cal R} + V_*(l,0,l)  \\
1, & \mbox{else.}
\end{array}
\right. \label{eq:dppolicy}
\end{equation}
for $i_0\in S$ and $1 \leq k \leq k_{max}$.

Thus, for a given paging RCL $f$, we have identified how to
compute a registration RCL $g$ to minimize $C(f,g)$.

\subsection{Optimal paging RCL for given registration RCL}

Suppose a registration RCL $g$ is fixed. In this subsection we
address the problem of finding a paging RCL $f$ to minimize
$C(f,g)$. For $i_0\in S$ and $0\leq k \leq k_{max}$, let
$w(i_0,k)$ denote the conditional probability distribution of the
state of the MS,  given that the most recent \report\ occurred $k$
time units earlier and the state at the time of the most recent
\report\ was $i_0$. Thus, $w(i_0,0)=\delta(i_0)$, and for larger
$k$ the $w$'s can be computed by the recursion:
\begin{displaymath}
w(i_0,k+1)=\Phi(w(i_0,k),g(i_0,k))
\end{displaymath}
The paging order vector $f(i_0,k)$ is simply the one to
be used when the MS must be paged $k$ time units after the
previous \report.  At such time the conditional distribution of
the state of the MS given the observations of the base station is
$w(i_0,k-1)P$.  Thus, the probability the MS is located in cell
$c$, just before the paging begins is given by
\begin{displaymath}
p(c|i_0,k)=\sum_{j\in S} \sum_{l\in c} w_j(i_0,k-1)p_{jl}
\end{displaymath}
Finally, $f(i_0,k)$ is the paging order vector for ordering the
cells $c$ according to decreasing values of the probabilities
$p(c|i_0,k).$

\subsection{Iterative optimization algorithm}
\label{algorithm}

In the previous subsections we described how to find an
optimal $g$ for given $f$ and vice versa.  This suggests an
iterative method for finding an individually optimal pair $(f,g)$.
The method works as follows.  Fix an arbitrary registration RCL
$g^0$. Then execute the following steps.
\begin{itemize}
\item
Find a paging       RCL $f^0$ to minimize $C(f^0,g^0)$
\item
Find a registration RCL $g_1$ to minimize $C(f^0,g^1)$,
\item
Find a paging       RCL $f^1$ to minimize $C(f^1,g^1)$, and so on.
\end{itemize}
Then $C(f^0,g^0) \geq C(f^0,g^1) \geq C(f^1,g^1) \geq C(f^1,g^2)
\geq \cdots $ Since there are only finitely many RCLs, it
must be that for some integer $d$, $C(f^d, g^d)=C(f^d,g^{d+1})$. By
construction, the paging RCL $f^d$ is optimal given the
registration RCL $g^d$. Similarly, $g^{d+1}$ is optimal given
$f^d$. However, since $C(f^d, g^d)=C(f^d,g^{d+1})$, it follows
that $g^d$ is also optimal given the registration RCL $f^d$.
Therefore, $(f^d,g^d)$ is an individually optimal pair of RCLs.

\section{EXAMPLES} \label{sec:examples}

Two examples are given in this section. Additional examples based on
random walk models are in the next section.

\subsection{Rectangular grid example}
\begin{figure}
\post{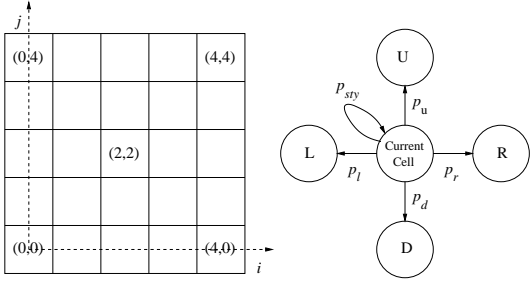}{7} \caption{Rectangular grid motion
model} \label{rect}
\end{figure}

Consider a rectangular grid topology, such that each cell has four neighbors.
The diagram to the left in Figure~\ref{rect} shows the finite
$i_{max} \times j_{max}$ rectangular grid topology.  To provide
the full complement of four neighbors to cells on the edges of the
grid, the region is wrapped into a torus. The torus
 can serve to approximate larger sets of cells.  Also, by the
symmetry of the torus, the functions $f(i_0,k),g(i_0,k)$ and
distributions $w(i_0,k)$ need be computed for only one value of
last reported cell $i_0$.
Each cell in Figure~\ref{rect} is represented by the index pair
$(i,j)$, where $i=0,1,\ldots,i_{max}-1$ is the index for the
horizontal axis, and $j=0,1,\ldots,j_{max}-1$ is the index for the
vertical axis.

\begin{figure}
\post{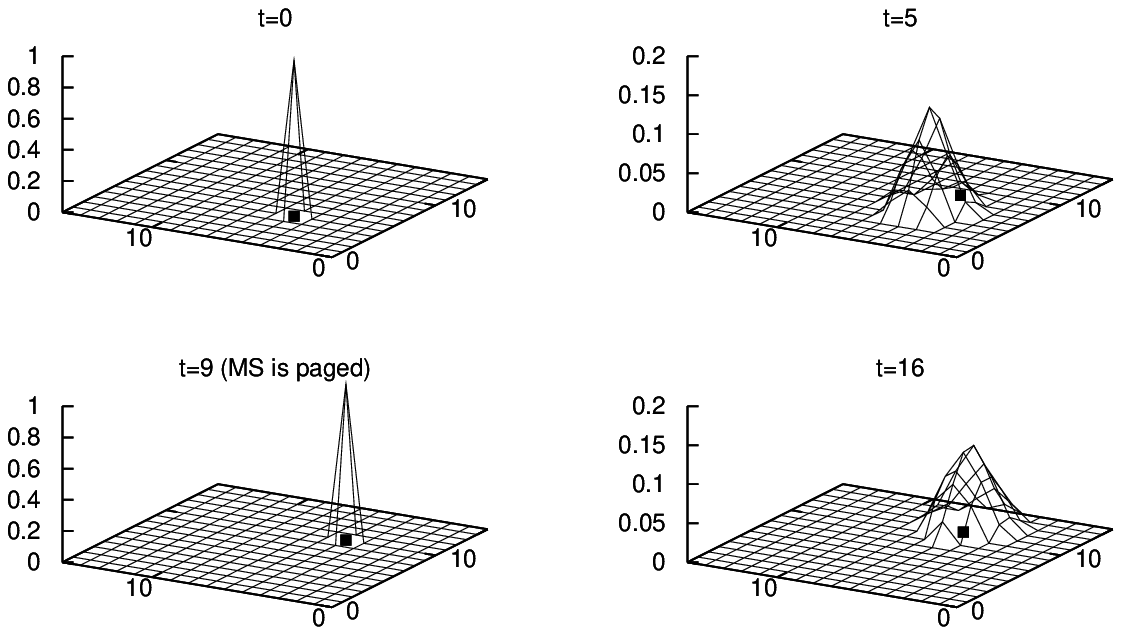}{8} \\
\post{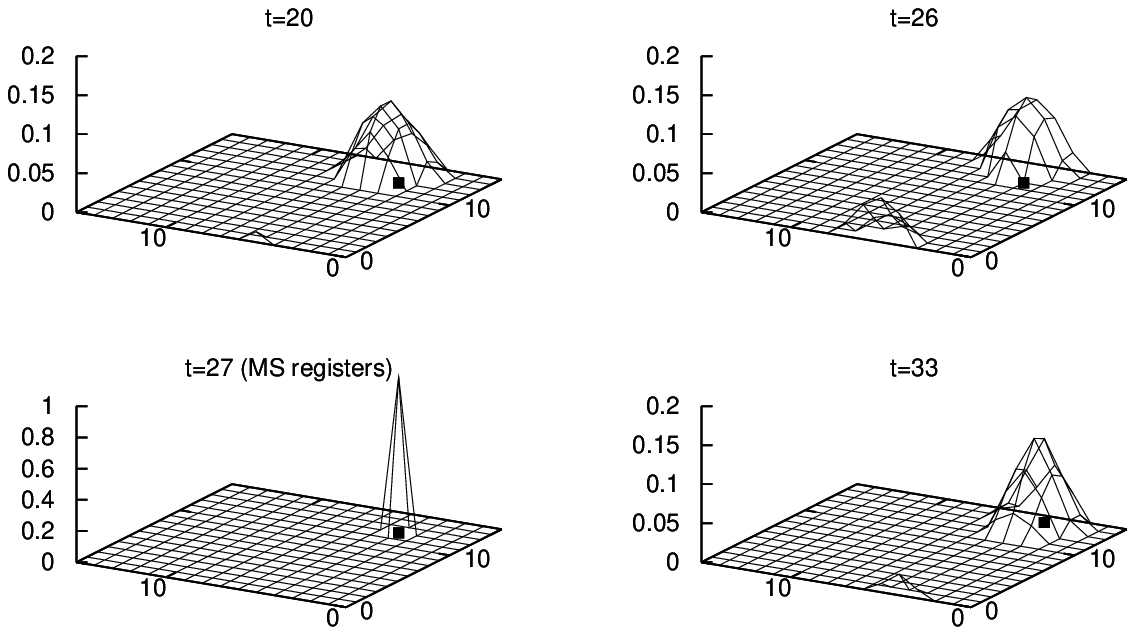}{8}
\caption{Evolution of the state $X(t)$ and the conditional
distribution of the state $w(t)$ for
the rectangular grid example.} \label{fig:rectmove}
\end{figure}

For simplicity, we assume that there is only one state per cell, so we can take
${\cal C}=S.$  For a numerical example, consider a $15 \times 15$ torus grid
 with motion parameters $p_{sty}=0.4$,
$p_{u}=p_{d}=p_{l}=0.1$, $p_{r}=0.3$, $x_0=(5,5)$ and other parameters
$\lambda_p=0.03$, ${\cal P}=1$, ${\cal R}=0.6$,
$\beta=0.9$, and $k_{max}=200$.  We numerically calculated an
individually optimal pair $(f,g)$ of RCLs.  A sample path of $X$ and $w$ generated
using those controls is indicated in Figure ~\ref{fig:rectmove}.  The figure shows for
selected times $t$ the state
$X(t)$, indicated by a small black square, and the conditional
state distribution $w(t)$, indicated as a moving bubble.  The
distribution $w(t)$ collapses to a single unit mass point at
$t=9$ due to a page and at $t=27$ due to a registration.
Roughly speaking, the MS registers when it is not where the
network expects it to be, given the last report received by the
network.   For instance, at time $t=26$ the MS is located at
the tail edge of the bubble, so the network has low accuracy
in guessing the MS location.  One time unit later, at t=27, the MS finds
itself so far from where the network thinks it should be that the
MS registers.

\subsection{Simple Example}  \label{sec:simp_example}

\begin{figure}
\post{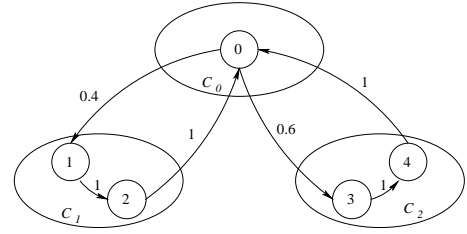}{6} \caption{Simple example}
\label{fig:suboptimal}
\end{figure}

The following is an example of a small network for which jointly
optimal paging and registration policies can be computed. The
example also affords a pair of individually optimal RCLs which
are not jointly optimal. The space structure of the example is shown in
Figure~\ref{fig:suboptimal}. $S=\{0,1,2,3,4\}$ and
${\cal C}=\{c_0,c_1,c_2\}$ with $c_0=\{0\}$, $c_1=\{1,2\}$,
$c_2=\{3,4\}$. From state 0, the MS transits to state 1 with
probability 0.4 and to state 3 with probability 0.6. The other possible transitions shown in the figure have probability 1. The
initial state is taken to be $0$.

\begin{figure}
\post{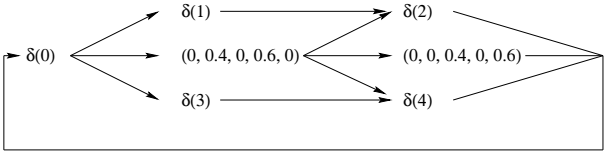}{8} \caption{Evolution of $w(t)$ for
the simple example.} \label{fig:trans}
\end{figure}

We first describe the jointly optimal pair of paging and
registration policies. We consider,
without loss of optimality, policies given by feedback control
laws ($\bar{f}$,$\bar{g}$) as described in Section
\ref{sec:joint}. Thus we take $u(t)=\bar{f}(w(t-1))$ and
$v(t)=\bar{g}(w(t-1))$. Due to the special structure of this
example, the process $w(t)$ takes values in a set of at most seven
states, and the possible transitions are shown in
Figure~\ref{fig:trans}. The dynamic programming problem for jointly optimal policies thus
reduces to a finite state problem. The optimal choice of the
mapping $\bar{f}$ is given by $\bar{f}^*(w)$, which pages states
in decreasing order of $wP$. It remains to find the optimal
registration policy mapping $\bar{g}$.

We claim that if $t~mod~3=0$ or $t~mod~3=2$, then it is optimal to
not register at time $t$. Indeed, if $t~mod~3=0$ then the network
already knows the MS is in state 0, so registration would cost
${\cal R}$ and provide no benefit. If $t~mod~3=2$, then the
network knows that the MS will be in state 0 at time $t+1$, which
is the next time of a potential page. Thus, again the registration
at time $t$ would cost ${\cal R}$ and provide no benefit. This
proves the claim.

Therefore, it remains to find the optimal registration vector
$v(t)$ to use when $t~mod~3=1$. Such vector is deterministic,
given by $\bar{g}(\delta(0))$. There are essentially only four
possible choices for $\bar{g}(\delta(0))$, as indicated in
Table~\ref{tab:regchoices}.

\begin{table}
\caption{Registration policies
$\bar{g}_A$, $\bar{g}_B$,$\bar{g}_C$, $\bar{g}_D$
for the simple example.}
\centering
\begin{tabular}{|c|c|c|c|}
\hline
Policy&$\bar{g}(\delta(0))$&$P \left[ R_1|P_1^c \right]$&$P \left[ N_2=2 |
P_1^c\cap P_2 \right]$\\\hline
A&$(0,0,0,0,0)$&0&0.4\\
B&$(0,1,0,0,0)$&0.4&0\\
C&$(0,0,0,1,0)$&0.6&0\\
D&$(0,1,0,1,0)$&1&0\\
\hline
\end{tabular}
\label{tab:regchoices}
\end{table}

The cost for any pair ($\bar{f}$,$\bar{g}$) is given by
\begin{eqnarray*}
&& C(\bar{f},\bar{g}) = \frac{{\cal R} \beta
(1-\lambda_p)P \left[ R_1|P_1^c \right] }{1-\beta^3}~~~~~~~~~~~~~~~~~~~~~~~~~~~ \\
&& + \frac{\lambda_p{\cal
P}(1.4\beta+\beta^2+\beta^2(1-\lambda_p)P \left[ N_2=2 | P_1^c \cap
P_2 \right] + \beta^{3})}{1-\beta^{3}}   
\end{eqnarray*}
Consulting Table~\ref{tab:regchoices} we thus find that ($\bar{f}^*$, $\bar{g}_A$) is
jointly optimal if ${\cal R} \geq \lambda_p{\cal P}\beta$, and
($\bar{f}^*$, $\bar{g}_B$) is jointly optimal if ${\cal R} \leq
\lambda_p{\cal P}\beta$.

For the remainder of this example we consider policies given by RCLs. 
Under the assumption that $0 < {\cal R} \leq \lambda_p {\cal P}
\beta$, the pair of mappings ($\bar{f}^*$, $\bar{g}_B$) is equivalent to a pair of RCLs, which
we denote by ($f_B$, $g_B$). Under $g_B$, the MS registers only
after entering state 1 and not being paged. The pair ($f_B$,$g_B$)
is jointly optimal, and hence it is also individually optimal.
Similarly, let ($f_C$,$g_C$) be RCLs corresponding to the feedback mappings
($\bar{f}^*$,$\bar{g}_C$). In particular, an MS using registration
RCL $g_C$ registers only after entering state $3$ and not being
paged.
\begin{prop}
The pair of RCLs ($f_C$,$g_C$) is individually optimal, but not jointly optimal.
\end{prop}

\begin{proof}
The paging RCL $f_C$ is optimal for the registration RCL $g_C$
because for $g_C$ fixed, it is equivalent to the optimal
feedback mapping $\bar{f}^*$. Suppose then that the MS uses the
paging RCL $f_C$. Note that if the MS does not report at time
$t=1$, and if it is paged at time $t=2$, the network will page
cell $c_1$ first. Hence, if the MS enters state 3 at time $t=1$ and
if it is not paged at $t=1$, then by registering for cost ${\cal R}$
it can avoid the two or more pages required at time $t=2$ in case
of a page at $t=2$. Since ${\cal R} \leq \lambda_p {\cal P} \beta$, it is
optimal to have the MS register at $t=1$ in this situation.
Thus $g_C$ is optimal for $f_C$, so the pair is individually optimal. However,
$C(f_C,g_C) > C(f_B,g_B)$, so that ($f_C, g_C$) is not jointly optimal.
\end{proof}

\section{Jointly Optimal Policies for Some Random Walk Models}  \label{sec:random_walk}

The structure of jointly optimal paging and registration policies are identified
in this section for three random walk models of motion.  The first is
a discrete state one-dimensional random walk, the second is for a symmetric
random walk in $\IR^d$ for any $d  \geq 1$, and the third is for a Gaussian
random walk in  $\IR^d$ for any $d  \geq 1$.

\subsection{Symmetric random walk in $\IZ$} \label{sec:random_walk_discrete}. 

Suppose the motion of the MS is modeled by a discrete-time random walk on an
infinite linear array of cells, such that the displacement of the walk at each step
has some probability distribution $b.$
Equivalently, $(X(t):t\geq 0)$ is a discrete
time Markov process on $\IZ$ with one-step transition
probability matrix $P$ given by $p_{ij}=b_{j-i}$.
For any probability distribution $w$, $wP=w*b$.
It is assumed that $b_i$ is a nonincreasing function of $|i|$,
or in other words, $b$ is symmetric about zero and unimodal.
In the general form of our model,  multiple states can correspond
to the same cell, but for this example, each integer state $i$ corresponds
to a distinct cell in which the MS can be paged.
So ${\cal C}=S=\IZ.$  It  is assumed that the network knows the initial
state $x_0$.

Due to the translation invariance of $P$ for this example, the update equations
of the dynamic program are translation invariant, and therefore the paging and
registration RCLs can also be taken to be
translation invariant.     Thus, we write the RCLs as $f=(f(k):k\geq 1)$
and $g=(g(k):k\geq 1)$.  These RCLs give the control decisions if the
last reported state is $i_0=0$, and hence for other values of $i_0$ by
translation in space.

It turns out that for this example, the optimal paging
policy is ping-pong type: cells are searched in an order of
increasing distance from the cell in which the previous report
occurred.  The optimal registration policy is a distance threshold
type: the mobile station registers whenever its distance from
the previous reporting point exceeds a threshold.
Specifically,  only RCLs of the following form need to
be considered.  The actions of the policies do not depend on the
time $k$ elapsed since last report, so the argument $k$ is suppressed.
For the paging policy we take the ping-pong policy, given by the RCL
$f^*=(0,1,-1,2,-2,3,-3,\ldots )$.  Thus, if the MS is to be
paged and if it was last reported to be at state $i_0$, then the
states are searched in the order $i_0,  i_0+1, i_0-1, i_0+2, i_0-2, \ldots$.
The registration policy is given by the RCL
$g^*_l=I_{\{ l \geq d_r ~or~ l\leq -d_l\}}$ where the two
distance thresholds $d_l, d_r \geq 1$  are such that either
$d_l=d_r$ or $d_l=d_r-1$.  
\begin{prop}  \label{prop:distthresh}
There is a choice of the distance thresholds $d_l$ and $d_r$
such that the ping-pong paging policy given by $f^*$
and the distance-threshold registration policy given by $g^*$
are jointly optimal.
\end{prop}
 
The related work of Madhow, Honig, and Steiglitz
\cite{MadhowHonigSteiglitz95} finds the
optimal registration policy assuming that the paging
policy is fixed to be the ping-pong policy.    Also, it is not
difficult to show that for the distance threshold registration
policy specified by $g^*$, the optimal paging policy is the ping-pong
paging policy.  However, a pair of individually optimal RCLs may
not be jointly optimal, as shown in the example of Section \ref{sec:simp_example}.

The remainder of this section is devoted to the proof of Proposition \ref{prop:distthresh}.
The following notation is standard in the theory of majorization \cite{MarshallOlkin79}.
Given $x= (x_1, x_2, \cdots, x_n)  \in \IR^n$,  let $x_{\downarrow} $ denote the nonincreasing rearrangement of
$x$.  That is, $x_{\downarrow} = (x_{[1]} , x_{[2]}, \cdots, x_{[n]})$, where  the coordinates
 $x_{[1]} , x_{[2]}, \cdots, x_{[n]}$ are equal to a rearrangement of
 $x_1, x_2, \cdots, x_n$, such that  $x_{[1]}  \geq x_{[2]} \geq \cdots \geq  x_{[n]}.$
 Given two vectors $x$ and $y$, we say that $y$ majorizes $x$, denoted by
 $x \prec y$, if the following conditions  hold:
\begin{eqnarray*}
 \sum_{i=1}^r x_{[i]}  &\leq & \sum_{i=1}^r y_{[i]}~~\mbox{for}~1\leq r \leq n-1 \\
  \sum_{i=1}^n x_{[i]}  & =  & \sum_{i=1}^n y_{[i]}    \label{eq.major_eq_n}
\end{eqnarray*}
Write $x \equiv y$ to denote  that both $x \prec y$ and $y \prec x$, meaning that $y$ is
a rearrangement of $x$.  The relation
$x \prec y$ can be defined in a similar fashion, in case $x$ and $y$ are nonnegative, summable functions defined on some countably infinite discrete set.   In such case,
$x_{[i]}$ denotes the $i^{th}$ coordinate, when the coordinates of $x$ are listed in a
nonincreasing order.  

Given a probability distribution $\mu$  on $\IZ$,  let $s(\mu)$ denote the mean
number of states that must be searched to find the MS, given that the MS has distribution
$\mu$ and the optimal search order for $\mu$ is used.  The
optimal search order is maximum likelihood search
\cite{Rose95}, under which states are searched in order of
decreasing probability.  Summation by parts yields
\begin{displaymath}
s(\mu)=\sum_{i=1}^\infty  i \mu_{[i]}  
=1+\sum_{r=1}^\infty (1-\sum_{i=1}^r \mu_{[i]}  ),
\end{displaymath}
which immediately implies the following lemma.
\begin{lemma}  \label{lemma:sorder}
If $\mu$ and $\nu$ are probability distributions such that
$\mu\prec \nu$, then $s(\mu)\geq s(\nu)$.
\end{lemma}

A function or probability distribution $\mu$ on $\IZ$ is said to be
{\em neat} if  $\mu_0 \geq \mu_1 \geq \mu_{-1} \geq \mu_2 \geq \mu_{-2} \geq \ldots$.

\begin{lemma} \label{lemma:neat}
If $\mu$ is a neat probability distribution, then the convolution
$\mu*b$ is neat.
\end{lemma}
\begin{proof}
For $i\geq 0,$ let $b^{(i)}$ denote the uniform probability
distribution over the interval of integers $[-i,i]$.
The conclusion is easy to verify in case $b$ has the form
$b^{(i)}$ for some $i$.   In general, $b$ is a convex combination
of such $b^{(i)}$'s, and then $\mu*b$ is a convex combination of the
functions $\mu*b^{(i)}$, using the same coefficients.  
Convex combinations of neat distributions are neat, so $\mu*b$ is indeed neat.
\end{proof}

\begin{lemma} \label{lemma:compact}
If $\mu$ and $\nu$ are probability distributions such that
$\mu \prec \nu$ and $\nu$ is neat, then $\mu*b \prec \nu *b $.
\end{lemma}
The proof of the Lemma \ref{lemma:compact} is placed in the appendix because the
proof is specific to the discrete state setting.    Lemma \ref{lemma:compact_cont} in
the next subsection is similar, and its proof shows the connection to
Riesz's rearrangement inequality.

Let $\mu$ be a probability distribution on $\IZ$ and let $0\leq\lambda <  1$.
Let ${\cal T}(\mu,\lambda)$ be the set of probability distributions
$\nu$ on $\IZ$ such that $(1-\lambda)\nu \leq \mu$, pointwise.  Intuitively,
such a $\nu$ is obtained from $\mu$ by trimming
away from $\mu$ probability mass $\lambda$ and renormalizing
the remaining mass.  The following lemma has an easy proof which
is left to the reader.  Roughly speaking, the lemma
means that given $\mu$ and $\lambda$, the
most maximal distribution in ${\cal T}(\mu, \lambda ),$ in the majorization order, is obtained by
trimming mass from the smallest $\mu_i$'s.
\begin{lemma} (Optimality of minimum likelihood trimming)
\label{lemma:trim}
There exists $\nu \in {\cal T}(\mu,\lambda)$ such
that for some $k\geq 1$,
\begin{displaymath}
(1-\lambda) \nu_{[j]} =\left\{
\begin{array}{cl}
\mu_{[j]}  & \mbox{if}~j<k \\
      0       & \mbox{if}~j>k. 
      \end{array} \right.
\end{displaymath}
Furthermore, for any other $\nu' \in {\cal T}(\mu,\lambda)$,
$\nu' \prec \nu$.
\end{lemma}


Let $f$ and $g$ be RCLs (possibly dependent on the elapsed
time $k$ since last report).  The cost $C(f,g)$ can be computed by
considering the process only up until the first time $\tau$ that
a report occurs (i.e. one reporting cycle).
Let $\alpha(k)=P[\tau=k]=\alpha_p(k)+\alpha_r(k)$, where
$\alpha_p(k)$ is the probability $\tau=k$ and the first report
is a page, and $\alpha_r(k)$ is the probability $\tau=k$ and the
first report is a registration.  Also let $w(k)$ denote the
conditional distribution of the MS given that no report occurs
up to time $k$ for the pair of RCLs $(f,g)$.  Then
\begin{eqnarray*}
C(f,g) & = & \frac{E\left[ \sum_{t=1}^\tau \beta^t\{{\cal P}I_{P_t}N_t
+{\cal R}I_{R_t}\}\right]}{1-E[\beta^\tau]}  \\
& = & \frac{ \sum_{k=1}^\infty \beta^k \{ {\cal P}
   \alpha_p(k)s(w(k-1)*b)+{\cal R}\alpha_r(k) \} }{1
     -\sum_{k=1}^\infty \beta^k\alpha(k) }
\end{eqnarray*}
Note that the cost depends entirely on the $\alpha$'s and on the
mean numbers of pages required, given by the terms $s(w(k-1)*b).$

\begin{lemma}  \label{lemma:page}
(Optimality of ping-pong paging $f^*$)
There exists a registration RCL $g^o$ so that $C(f^*,g^o)\leq C(f,g)$.
\end{lemma}
\begin{proof}
Take the registration RCL $g^o$ to be of distance threshold
type with time varying thresholds and possibly with
randomization at the left threshold if the thresholds are
equal, or at the right threshold if the right threshold is
one larger than the left threshold.  More precisely, for
fixed $k$: all the values
$g_l^o(k)$ are binary except for possibly one value of $l$,
and $1-g^o(k)$ is neat.  Select the thresholds and randomization
parameter so that the $\alpha$'s, $\alpha_p$'s, and
$\alpha_r$'s are the same for the pair $(f^*,g^o)$ as for the
originally given pair $(f,g)$.  

Let $(w^o(k): k\geq 0)$ and $\tau^o$ be defined for $(f^*,g^o)$
just as $(w(k): k\geq 0)$ and $\tau$ are defined for $(f,g).$
To complete the proof of the lemma it remains to
show that $s(w^o(k-1)*b)\leq s(w(k-1)*b)$ for $k\geq 1$.
The sequences $w$ and $w^o$ are updated in similar ways, by
Lemma \ref{lemma:w_rep}:
\begin{eqnarray*}
w(k) & = & \Phi (w(k-1),g(k))~~~~1\leq k \leq \tau-1 \\
w^o(k) &= & \Phi (w^o(k-1),g^o(k) )~~~1\leq k \leq \tau-1 \\
\end{eqnarray*}
By the definition of $\Phi$, this means the distribution $w(k)$ is
obtained by first forming the convolution $w(k-1)*b$, removing a
fraction $g_l(k)$ of the mass at each location $l$, and renormalizing
to obtain a probability distribution.  The RLC $g^o$ trims mass in a
minimum likelihood fashion.  Thus, Lemmas \ref{lemma:neat},
\ref{lemma:compact}, and \ref{lemma:trim} to show
by induction that for all $k\geq 1$:
$w^o(k)$ is neat,  $w(k)\prec w^o(k)$, and
$w(k-1)*b \prec  w^o (k-1)*b$.
Thus by Lemma \ref{lemma:sorder},
$s(w^o(k-1)*b)\leq s(w(k-1)*b)$, completing the
proof of Lemma \ref{lemma:page}.
\end{proof}

{\em Proof of Proposition \ref{prop:distthresh}. }
In view of Lemma \ref{lemma:page}, it remains to show
that if the ping-pong paging policy specified by $f^*$ is used, then 
for some choice of fixed distance thresholds $d_l$ and $d_r$, the
registration policy specified by $g^*$ is optimal.  This can be done by
examining a dynamic program for the optimal registration
policy, under the assumption that the RCL $f^*$ is used.
Let $V_n(j)$ denote the mean discounted cost for
$n$ time steps to go, given that the mobile is located directed
distance $j$ from its last reported state.  Then
\begin{eqnarray*}
V_{n+1}(j) & = & \beta\sum_{l\in \IZ}b_{j-l}
[ \lambda_p( {\cal P} f_l^*+V_n(0)) \\
 && + (1-\lambda_p)\min\{V_n(l),{\cal R}+V_n(0)\}]
\end{eqnarray*}
By a contraction property of these dynamic programming
equations, the limit $V_*=\lim_{n\rightarrow \infty} V_n$
exists.   Argument by induction yields that the functions
$-V_n$ are neat, and hence that $-V_*$ is neat.  
By the dynamic programming principle, an optimal registration
policy  is given by the RCL $g^*$ specified by:
\begin{displaymath}
g^*_l=\left\{ 
\begin{array}{cl}
1 & \mbox{if}~V_*(l)\geq {\cal R}+V_*(0) \\
0 & \mbox{else}
\end{array} \right.
\end{displaymath}
Since $-V_*$ is neat, the optimal registration RCL $g^*$
has the required threshold type.  
Proposition \ref{prop:distthresh} is proved.
\hfill\rule{.75em}{.75em}\bigskip

\subsection{Symmetric random walk in $\IR^d$} \label{sect:dsymmetric}

To extend Proposition \ref{prop:distthresh} to more than one dimension,
we consider a continuous state mobility model, with
$S={\cal C}=\IR^d$, for an integer $d \geq 1$.    Of course in practice we expect
$d \leq 3$.
A function on $\IR^d$ is said to be {\em symmetric nonincreasing} if it can be expressed as
$\phi(|x|)$, for some nonincreasing function $\phi$ on $\IR_+$, where $|x|$ denotes the
usual Euclidean norm of $x$.
Let $x_o \in \IR^d$, and let $b$ be a symmetric nonincreasing probability density function (pdf)
on $\IR^d$.  
The location of the MS at time $t$ is assumed to be given by $X(t)=x_o+ \sum_{s=1}^t B_s$,
where the initial state $x_o$ is known to the network, and the random variables
$B_1, B_2, \ldots$ are independent, with each having pdf $b$.

Let  ${\cal L}^d(A)$ denote the volume (i.e. Lebesgue measure) of a Borel set
$A \subset \IR$.   A {\em paging order function}  $r=(r_x:x\in \IR^d)$ is a nonnegative
function on $\IR^d$ such that ${\cal L}^d(\{ x : r_x \leq \gamma \})=\gamma$ for all $\gamma \geq 0.$ 
Thus, as $\gamma$ increases, the volume of the set  $\{ x : r_x \leq \gamma \}$ increases at
unit rate.  Imagine the set  $\{ x : r_x \leq \gamma \}$ increasing as $\gamma$ increases,
until the MS is in the set.   If the MS is located at $\bar{x}$ and is paged according to the paging order
function $r$,  then $r_{\bar{x}}$ denotes the volume of the set searched to find $\bar{x}$.
So the paging cost is ${\cal P}r_{\bar{x}}$, where $\cal P$ is the cost of paging per unit volume
searched.  An example of a paging order is {\em increasing distance search, starting at $x_o$},
which corresponds to letting  $r_x$ be the volume of a ball of  radius $|x-x_o|$ in $\IR^d.$
As in the finite state model,  assume the cost of a registration is $\cal R$.

Paging and registration policies $u$ and $v$ can be defined for this model just as they
were for the finite state model, with paging order functions playing the
role of paging order vectors.   Thus, for each $t \geq 1$, $u(t)=(u_x(t): x\in \IR^d)$ is a paging order function, and $v(t)=(v_x(t): x\in \IR^d)$ is a $[0,1]$-valued function.
In addition, translation invariant RCLs $f$ and $g$ can be defined as they were for the
one-dimensional network model, and they determine policies $u$ and $v$ as follows.
If the location of the most recent report was $x_o$,
then  $u_x(t) = f_{x-x_o}$ and $v_x(t)=g_{x-x_o}$.  
Let $f^*$ be the RCL for increasing distance search paging: $f^*_x$ is the volume of the radius $|x|$ ball in $\IR^d$.  Let $g^*$ be the RCL for the distance threshold registration policy
with some threshold $\eta$:  $g^*_x=I_{\{|x| \geq \eta\}}$.

\begin{prop}  \label{prop:distthresh_cont}
There is a choice of the distance threshold $\eta$
such that $f^*$ and $g^*$ are jointly optimal.
\end{prop}

The proof of Proposition \ref{prop:distthresh} can be used for the proof of Proposition
\ref{prop:distthresh_cont}, with  symmetric nonincreasing functions on $\IR^d$
replacing neat probability distributions on $\IZ.$   A suitable variation of 
Lemma  \ref{lemma:compact}  must be established, and
we will show that this can be done by applying Riesz's rearrangement  inequality.
To get started, we introduce some notation from the theory of rearrangements of functions
(similar to the notation in \cite{LiebLoss01}.)
If  $A$ is a Borel subset of $\IR^d$
with ${\cal L}^d(A) < \infty$, then the {\em symmetric rearrangement of $A$}, denoted by
$A^{\sigma}$, is the open ball in $\IR^d$ centered at 0 such that
${\cal L}^d(A) ={\cal L}^d(A^{\sigma})$.  Given an integrable, nonnegative
 function $h$ on $\IR^d$, its {\em symmetric nonincreasing rearrangement}, $h^{\sigma}$, is
defined by
$$
h^{\sigma}(x) = \int_0^{\infty} I_{ \{h > t \}^{\sigma}  }  dt
$$
Let $h_1*h_2$ denote the convolution of functions $h_1$ and $h_2$, and let
$(h_1,h_2)=\int_{\IR^d} h_1h_2 ~dx$.
A proof of the following celebrated inequality is given in \cite{LiebLoss01}.
%
%
\begin{lemma}{\em F. Riesz's rearrangement inequality\cite{Riesz30}})  If $h_1$, $h_2$, and $h_3$ are nonnegative functions on $\IR^d$, then  \\$( h_1 , h_2*h_3)    \leq 
( h^\sigma_1 , h_2^\sigma * h_3^\sigma ).$
\end{lemma}

Given two probability densities on $\IR^d$, $\nu$ majorizes
$\mu$, written $\mu \prec \nu$, if
$$
\int_{|x| \leq \rho} \mu^{\sigma} ~ dx \leq \int_{|x| \leq \rho} \nu^{\sigma} ~ dx~~\mbox{for all}~\rho > 0.
$$
Equivalently,   $\mu \prec \nu$ if, for any Borel set $F \subset \IR^d$, there is another
Borel set $F' \subset \IR^d$ with ${\cal L}^d(F)={\cal L}^d(F'),$ such that
$$
\int_F \mu ~ dx \leq \int_{F' } \nu ~ dx.
$$
If $\mu \prec \nu$, then $(\mu^\sigma , h) \leq (\nu^\sigma , h)$,
for any symmetric  nonincreasing function $h.$  (To see this, use the fact that
such an $h$ is a convex combination of indicator functions of balls centered at zero.)

\begin{lemma} \label{lemma:compact_cont}
If $\mu$ and $\nu$ are probability densities such that
$\mu \prec \nu,$ and if $\nu$ is symmetric nonincreasing , then $\mu*b \prec \nu *b $.
\end{lemma}
\begin{proof}
Let $F$ be an arbitrary Borel subset of $\IR^d.$ 
Let $h_1=\mu$, $h_2=I_F$, and $h_3=b$.  Then
$h_1^\sigma=\mu^\sigma$, $h_2=I_{F^\sigma}$, $h_3=b$, and
Riesz's rearrangement inequality yields $(\mu, I_F * b ) \leq (\mu^\sigma, I_{F^\sigma}*b).$
Since $\mu \prec \nu=\nu^\sigma$ and $I_{F^\sigma}*b$ is symmetric nonincreasing,
$(\mu^\sigma , I_{F^\sigma}*b) \leq (\nu,  I_{F^\sigma}*b )$.   Combining yields
$(\mu, I_F * b ) \leq (\nu,  I_{F^\sigma}*b )$, or, equivalently by the symmetry of $b$,
$(\mu*b, I_F  ) \leq (\nu*b,  I_{F^\sigma} )$.   That is,
$$
\int_F  \mu*b ~ dx \leq \int_{F^\sigma } \nu*b ~ dx.
$$
Since $F$ was an arbitrary Borel subset of $\IR^d$ and ${\cal L}^d(F)={\cal L}^d(F^\sigma  )$,
$\mu*b  \prec \nu*b $.
\end{proof}

{\em Proof of Proposition \ref{prop:distthresh_cont} }
Proposition \ref{prop:distthresh_cont}  follows from  Lemma \ref{lemma:compact_cont},
and the same arguments used to prove Proposition \ref{prop:distthresh_cont}. 
The details are left to the reader. 
\hfill\rule{.75em}{.75em}\bigskip

\subsection{Gaussian random walk in $\IR^d$}

Consider the following variation of the model of Section  \ref{sect:dsymmetric}.
Let $X(t)=x_o+ \sum_{s=1}^t B_s$, where the random variables $B_s$ are independent with
a $d$-dimensional Gaussian density with mean vector $m$ and covariance matrix $\Sigma$. 
Given a vector $y$ let $|y|_\Sigma = (y^* \Sigma^{-1} y)^{-1/2}.$
Proposition \ref{prop:distthresh_cont}   can be applied to the process with initial state
$\Sigma^{-1/2} x_o$ and increments $\widetilde{B}_i = \Sigma^{-1/2} (B_i-m).$  
Suppose the time of the last report was $t_o$ and the location at that time was $x_o$,
and suppose the MS just jumped to a new state at time $t$.  Let $\bar{x}(t)=x_o+(t-t_o)m$.
If the MS must be paged at time $t$, the optimal paging policy is to page according to
expanding  ellipses of the form  $\{ x : |x- \bar{x}(t) |_\Sigma   \leq \rho\}.$  If the MS is not
paged at time $t$, the optimal registration policy is for the MS to register if
$ |X(t) -  \bar{x}(t) |_\Sigma  \geq \eta$, for a suitable threshold $\eta$.

A continuous time version of this result can also be established, for which the motion
of the MS is modeled as a $d$-dimensional Brownian motion with drift vector $m$ and
infinitesimal covariance matrix $\Sigma$.

\section{Conclusions}
\label{sec:conclusion}

There are many avenues for future research in the area of paging and
registration.  This paper shows how the joint paging and registration optimization
problem can be formulated as a dynamic programming problem with
partially observed states. In addition, an iterative method is proposed,
involving dynamic programming with a finite state space, in order to find
individually optimal pairs of RCLs.  While an example shows that, in principle,
the individually optimal pairs need not be jointly optimal, no bounds
are given on how far from optimal the individually optimal pairs can be.
Furthermore, even the problem of finding individually optimal RCLs
may be computationally prohibitive, so it may be fruitful to apply
approximation methods such as neurodynamic programming
~\cite{BertsekasTsitsiklis96}. This becomes especially true if the model is
extended to handle additional features of real world paging and registration
models, such as the use of parallel paging,  overlapping registration regions,
congestion and queueing of paging requests for different MSs, positive probabilties
of missed pages, more complex motion models, estimation of motion models,
and so on. 

This paper shows that  jointly optimal paging and registration policies
for symmetric or Gaussian random walk models are given by
nearest-location-first paging policies and distance threshold  registration
policies.  It remains to be seen whether these policies are good ones,
even if no longer optimal, when the assumptions of the model are violated.
It also remains to be seen if jointly optimal policies can be identified
for other subclasses of motion models.

We found that majorization theory,  and, in particular, Riesz's rearrangement inequality,
are tools well suited for the study of a certain search algorithms with feedback.  These
tools may be more widely useful for addressing search or distributed sensing
problems.

\appendix
\section*{Appendix A: On $\sigma$-algebra notation}
Some basic definitions involving $\sigma$-algebras are collected in this appendix. In this paper
the network only observes random
variables with finite numbers of possible outcomes, so that emphasis is given to conditioning with
respect to finite $\sigma$-algebras.

The collections of random variables considered in this paper are
defined on some underlying probability space. A probability space
is a triple $(\Omega, {\cal F}, P)$, such that $\Omega$ is the set
of all possible outcomes, ${\cal F}$ is a $\sigma$-algebra of
subsets of $\Omega$  (so $\emptyset \in {\cal F}$ and $\cal F$ is
closed under complements and countable intersections) and $P$ is a
probability measure, mapping each element of $\cal F$ to the
interval $[0,1]$. The sets in $\cal F$ are called events. A random
variable $X$ is a function on $\Omega$ which is $\cal F$ measurable,
meaning that $\cal F$ contains all sets of the form $\{\omega :
X(\omega)\leq c\}$. In the remainder of this section, $\cal N$
denotes a $\sigma$-algebra that is a subset of $\cal F$.
Intuitively, $\cal N$ models the information available from some
measurement: one can think of $\cal N$ as the set of events that
can be determined to be true or false by the measurement. A random
variable $Y$ is said to be $\cal N$ measurable if $\cal N$
contains all sets of the form $\{\omega:Y(\omega)\leq c\}$.
Intuitively, $Y$ is $\cal N$ measurable if the information represented
by $\cal N$ determines $Y$.

An atom $B$ of $\cal N$ is a set $B\in {\cal N}$ such that if
$A\subset B$ and $A \in {\cal N}$ then either $A=\emptyset$ or
$A=B$. Note that if $C\in {\cal N}$ and $B$ is an atom of $\cal N$,
then either $B\subset C$ or $B\subset C^c$.  If $\cal N$ is finite
(has finite cardinality) then there is a finite set of atoms
$B_1,\ldots,B_m$ in $\cal N$  such that each element of $\cal N$
is either $\emptyset$ or the union of one or more of the atoms.

Given a random variable $X$ with finite mean, one can define
$E[X|{\cal N}]$ in a natural way.  It is an $\cal N$ measurable
random variable such that $E[XZ]=E[  E[X|{\cal N}]Z]$ for any
bounded, $\cal N$ measurable random variable $Z$. In particular, if
$A$ is an atom in $\cal N$, then $E[X|{\cal N}]$ is equal to
$E[XI_A]/P[A]$ on the set $A$. (Any two versions of $E[X|\cal N]$
are equal with probability one.)

Given a random variable $Y$, we write $\sigma(Y)$ as the smallest
$\sigma$-algebra containing all sets of the form $\{\omega\in
\Omega:Y(\omega)\leq c\}$. The notation $E[X|Y]$ is equivalent to
$E[X|\sigma(Y)]$. In case $Y$ is a random variable with a finite
number of possible outcomes $\{y_1,\ldots , y_m\}$, the
$\sigma$-algebra $\sigma(Y)$ is finite with atoms $B_i= \{ \omega  : Y(
\omega ) =y_i \}$, $1\leq i \leq m$.  Furthermore, given a random
variable $X$ with finite mean, $E[X|Y]$ is the function on $\Omega$ which is equal
to $\frac{E[XI_{B_i}]}{P[B_i]}$ on $B_i$ for $1\leq i
\leq m$.

\section*{Appendix B: Proof of Lemma \ref{lemma:w_rep} }
Since all the random variables generating ${\cal N}_{t+1}$
have only finitely many possible values, the $\sigma$-algebra
${\cal N}_{t+1}$ is finite.
Both sides of (\ref{eq:w_rep}) are ${\cal N}_{t+1}$ measurable,
so both sides are constant on each atom of ${\cal N}_{t+1}$.
Thus, if $A$ denotes an atom of ${\cal N}_{t+1}$, each side
of (\ref{eq:w_rep}) can be viewed as a function of $A$,
and it must be shown that the equality holds for all such
$A$.  Below we shall write $w_l(t+1,A)$ for the
value of $w_l(t+1)$ on the atom $A$. 

Since $P_{t+1}\cup R_{t+1} \in {\cal N}_{t+1}$, it follows that
either $A\subset P_{t+1}\cup R_{t+1}$ or $A\subset P^c_{t+1}\cap R^c_{t+1}$.
If $A\subset P_{t+1}\cup R_{t+1}$ then $X(t+1)$ is determined by $A$, 
and $w(t+1,A)=\delta(X(t+1)),$
so that (\ref{eq:w_rep}) holds on $A$.  So for the remainder of the
proof, assume that $A\subset P^c_{t+1}\cap R^c_{t+1}$. 

It follows that $A$ can be expressed as
$A = \widehat{A} \cap P^c_{t+1}\cap R^c_{t+1}$ for some atom $\widehat{A}$ of
${\cal N}_t$.  Thus for any state $l$
\begin{equation}  \label{eq:T_ratio}
w_l(t+1,A)=P[X(t+1)=l|A]=
\frac{T_l}{\sum_{l'\in S} T_l'}.
\end{equation}
where, 
letting $w_j(t,\widehat{A})$ denote the value of $w_j(t)$ on $\widehat{A}$ and
$v_l(t+1,\widehat{A})$ denote the value of $\widehat{v}_l(t+1)$ on $\widehat{A}$,
\begin{eqnarray*}
T_l & = & P[ R_{t+1}^c \cap \{ X(t+1)=l \} | \widehat{A} \cap P_{t+1}^c ] \\
& = & E[ (1-v_l(t+1))I_{\{X(t+1)=l\}} | \widehat{A} \cap P_{t+1}^c ]  \\
& = & E[ (1-v_l(t+1))I_{\{X(t+1)=l\}} | \widehat{A}  ]  \\
& = & P[ X(t+1)=l  | \widehat{A}  ] (1-\widehat{v}_l(t+1,\widehat{A} ) ) \\
& = & \left(  \sum_j w_j(t,\widehat{A})q_{jl}  \right)   (1-\widehat{v}_l(t+1,\widehat{A} ) ).
\end{eqnarray*}
Therefore
\begin{displaymath}
w_l(t+1,A)=\Phi_l(w(t,\widehat{A}),\widehat{v}(t+1,\widehat{A})),
\end{displaymath}
for any atom $A$ of ${\cal N}_{t+1}$ with $A \subset P^c_{t+1}\cap R^c_{t+1}$.
Lemma \ref{lemma:w_rep} is proved.

\section*{Appendix C: Proof of Lemma \ref{lemma:compact} } 

Lemma \ref{lemma:compact} is proved following the statement and proof
of three lemmas.

\begin{lemma}  \label{lemma:c0}
Consider two monotone sequences of some finite length $n$:
$a_1 \geq a_2 \geq \ldots \geq a_n=0 $ and
$0=b_1\leq b_2 \leq ... \leq b_n.$
Let $c_i=a_i+b_i$ for $1\leq i \leq n$, let
$d_i=a_i+b_{i+1}$ for $1 \leq i\leq n-1$, and let $d_n=0$.
Then $c \prec d$.
\end{lemma}
\begin{proof}
Note that $d_i\geq c_i$ and $d_i \geq c_{i+1}$ for $1\leq i \leq n-1$,
and the sum of the $c$'s is equal to the sum of the $d$'s .
  Therefore, for any subset
$A$ of $\{1,2,\ldots,n\}$, there is another subset $A'$ with
$|A|=|A'|$ such that  $\sum_{i\in A} c_i \leq \sum_{i\in A'} d_i$.
That proves the lemma.
\end{proof}

\begin{lemma} \label{lemma:c1}
Let $r$ and $L$ be positive integers.  Consider the convolution $F*G$
of two binary valued functions on $\IZ$, such that the support of $F$
has cardinality $r$, and the support of $G$ is a set of $L$
consecutive integers.   Then the convolution is maximal
 in the majorization order,  if the support of $F$
is a set of $r$ consecutive integers.
\end{lemma}
\begin{proof}
Suppose without loss of generality that $G=I_{\{ 0\leq i \leq L-1\} }$.
If the support of $F$ is not an interval of integers, let $j_{max}$ be the
largest integer in the support of $F$ and let $j_0$ be the smallest
integer such that the support of $F$ contains the interval of integers
$[j_o,j_{\max}]$.    Then $F=F^a+F^b$, such that $F^a_i=0$
for $i\geq j_0-1$ and the support of $F^b$ is the interval of
integers $[j_0,j_{\max}]$.  Let $F'$ be the new function defined
by $F'_i=F^a_i+F^b_{i+1}$.  The graph of $F'$ is obtained by sliding
the rightmost portion of the graph of $F$ to the left one unit.

We claim that $F*G\prec F'*G$.   To see this, note that
$F*G=F^a*G + F^b*G$. 
The idea of the proof is to focus on the interval of integers
$I=[j_0-1,j_0+r-2]$ and appeal to Lemma \ref{lemma:c0}.
The function $F^a*G$ is nonincreasing on $I$, it takes
value zero at the right endpoint of $I$, and it is also
zero everywhere to the right of $I$.  The function
$F^b*G$ is nondecreasing on $I$, it takes value zero at the
left endpoint of $I$, and it is also zero everywhere to the
left of $I$.  The convolution $F'*G$ is the same as $F*G$
except the second function $F^b*G$ is shifted one unit
to the right.  Lemma \ref{lemma:c0} thus implies that
$F*G \prec F'*G$. 
  This procedure can be repeated until $F$ is reduced to
a function with support being a set of $r$ consecutive integers. 
The lemma is proved.
\end{proof}

\begin{lemma}  \label{lemma:c2}
Let $r\geq 1$ and consider the convolution $F*b$ such that
$F$ is a binary valued function on the integers with support
of cardinality $r$.   Then the convolution is maximal  in the majorization order
if the support of $F$ consists of $r$ consecutive integers.
\end{lemma}
\begin{proof}
For $i\geq 0$, let $b^{(i)}$ denote the uniform probability
distribution on the interval $[-i,i]$, of $L=2i+1$ integers.
The lemma is true if $b=b^{(i)}$ for some $i$ by Lemma \ref{lemma:c1}.
Let $F^*$ denote the unique neat binary valued function with support
of cardinality $r$.   Note that  $F^* *b^{(i)}$ is neat for all $i \geq 0$
because both $b^{(i)}$ and $F^*$ are neat.
In general,  $b$ can be written as
$b=\sum_{i=0}^\infty \lambda_i b^{(i)}$ for some probability distribution
$\lambda$ on $\IZ_+$.   Therefore, for any binary $F$
with support of cardinality $r,$
\begin{eqnarray*}
b*F &   =   & \sum_i \lambda_i( b^{(i)}*F )
    \stackrel{(a)}{\prec}  \sum_i \lambda_i   (b^{(i)}*F)_{\downarrow}  \\
    & \stackrel{(b)}{\prec} &  \sum_i \lambda_i   (b^{(i)}*F^*)_{\downarrow}  
  = (b*F^*)_{\downarrow}  \equiv b*F^*.
\end{eqnarray*}
Here (a) follows from the fact that
taking nondecreasing rearrangements of  probability distributions
before adding them increases the sum in the majorization order, and
(b) follows from Lemma  \ref{lemma:c1}.
\end{proof}

{\em Proof of Lemma \ref{lemma:compact} }
Fix $r\geq 1$, let $F$ range over all binary valued functions
on $\IZ$ with support of cardinality $r$, and let $F^*$ denote the unique
choice of $F$ that is neat.   Use ``$(\mu, \nu )$" to denote inner
products. 
\begin{eqnarray*}
\sum_{i=1}^r  (\mu*b)_{[i]} & = &    \max_F ( \mu *b,  F )
        ~=~ \max_F ( \mu   , b*F ) \nonumber    \\
& \stackrel{(a)}{\leq} & \max_F ( \mu_{\downarrow}, (b*F)_{\downarrow} )  \\
&  \stackrel{(b)}{\leq }  &  ( \mu_{\downarrow}, (b*F^*)_{\downarrow} ) \\
& \leq &  ( \nu_{\downarrow}, (b*F^*)_{\downarrow} )  ~\stackrel{(c)}{=}~ ( \nu,b*F^*)  \\
& = & ( \nu*b,F^*)   ~\stackrel{(d)}{=}~  \sum_{i=1}^r  (\nu*b)_{[i]} 
\end{eqnarray*}
Here, (a) follows from the fact that rearranging each of two distributions in nonincreasing
order increases their inner product, (b) follows from Lemma \ref{lemma:c2} and the
monotonicity of $\mu_{\downarrow}$, (c) follows from the
fact that both $\nu$ and $b*F^*$ are neat, so their innner
product is the same as the inner product of their rearranged probability
distributions, and (d) follows from the fact that $\nu*b$ is neat.
\hfill\rule{.75em}{.75em}\bigskip

\section*{Acknowledgment}
The authors are grateful for useful discussions with
Rong-Rong Chen and Richard Sowers.


\begin{thebibliography}{1}

\bibitem{Aky96}
I.F. Akyildiz, J.S.M. Ho, and Y.B Lin, ``Movement-based location
update and selective paging for PCS networks,'' {\em IEEE/ACM
Transactions on Networking}, vol.~4, pp.~629--638, August 1996.

\bibitem{AnjumTassiulasShayman02}
F. Anjum, L. Tassiulas , and M. Shayman " Optimal paging for mobile location tracking ,"
{\em Advances in Performance Analysis ,} vol.  3. pp. 153-178, March 2002. 

\bibitem{Bar93}
A. Bar-Noy and I. Kessler, ``Tracking mobile users in wireless
communication  networks,'' {\em Proc.~IEEE Infocom}, pp.~1232--1239, 1993.

\bibitem{Bar94}
A. Bar-Noy, I. Kessler, and M. Sidi, ``Mobile users: To update or
not to update?'' {\em IEEE/ACM Transactions on Networking},
vol.~4, pp.~629--638, August 1996.

\bibitem{Bertsekas95a}
D. P. Bertsekas, {\em Dynamic Programming and Optimal Control,
Vol.~I}.
\newblock Athena Scientific, Belmont, MA. 1995.

\bibitem{Bertsekas95b}
D. P. Bertsekas, {\em Dynamic Programming and Optimal Control, Vol.~II}.

\newblock Athena Scientific, Belmont, MA. 1995.

\bibitem{BertsekasTsitsiklis96}
D. P. Bertsekas and J. N. Tsitsiklis, {\em Neuro-Dynamic Programming},
\newblock Athena Scientific, Belmont, MA. 1996.

\bibitem{Cachin97}
C. Cachin,  {\em Entropy Measures and Unconditional Security in Cryptography,}
PhD thesis, Swiss Federal Institute of Technology ZŸrich, 1997.

\bibitem{DeSantisGaggiaVaccaro01}
A. De Santis, A.G. Gaggia, and U. Vaccaro,
``Bounds on entropy in a guessing game," {\em IEEE Trans. Information Theory,}
vol.  47, pp. 468 - 473, 2001.



\bibitem{Hajek02}
B. Hajek, ``Jointly optimal paging and registration for a symmetric random walk,"
{\em IEEE Information Theory Workshop,}  Bangalore, India, pp. 20-23, October 2002.

\bibitem{HajekMitzelYang03}
B. Hajek, K. Mitzel, and S. Yang, "Paging and registration in cellular networks: jointly optimal policies and an iterative algorithm," {\em IEEE INFOCOM 2003,}  March  30-April 3, pp. 524 - 532, 2003.

\bibitem{Hass}
B. Krishnamachari, R.-H. Gau, S. B. Wicker, and Z. J. Haas, "Optimal sequential paging in cellular networks,"  {\em ACM Wireless Networks,} vol.~10, no. 2, pp. 121-131, March 2004. 

\bibitem{LiebLoss01}
E.H. Lieb and M. Loss, {\em Analysis}, second edition,  American Mathematical Society, Providence, 2001.

\bibitem{MadhavapeddyBasuRoberts95}
S. Madhavapeddy, K. Basu, and A. Roberts, ``Adaptive paging
algorithms for cellular systems,'' {\em Proc.~45th
IEEE Vehicular Technology Conference}, vol.~2, pp.~976--980, 1995.

\bibitem{MadhowHonigSteiglitz95}
U. Madhow, M. L. Honig, and K. Steiglitz, ``Optimization of
wireless resources for personal communications mobility
tracking,'' {\em IEEE/ACM Transactions on Networking}, vol.~3,
pp.~698--707, December 1995.

\bibitem{MarshallOlkin79}
A.W. Marshall and I. Olkin, {\em Inequalities: Theory of Majorization and Its Applications},
Academic Press, New York, 1979.

\bibitem{Massey94}
J.L. Massey,  ``Guessing and entropy,"    {\em Prof. 2004 IEEE International Symposium
on Information Theory,}  p. 204, Trondheim, Norway, 1994.

\bibitem{Riesz30}
F. Riesz, ``Sur une in\'egalit\'e int\'egrale. {\em J. London Mathematical Society},
vol. 5, pp. 162-168, 1930.

\bibitem{Rose95}
C. Rose and R. Yates, ``Minimizing the average cost of paging
under delay constraints,'' {\em Wireless Networks}, vol.~1,
pp.~211--219, 1995.

\bibitem{Rose96}
C. Rose, ``Minimizing the average cost of paging and registration:
A timer-based method,'' {\em Wireless Networks}, vol.~2,
pp.~109--116, 1996.

\bibitem{Rose99}
C. Rose, ``State-based paging/registration:a greedy technique,''
{\em IEEE Transactions on Vehicular Technology}, vol.~48,
pp.~166--173, January 1999.

\bibitem{Aky01}
W. Wang, I. F. Akyildiz, and G. L. Stuber, ``Effective paging
schemes with delay bounds as QoS constraints in wireless
systems,'' {\em Wireless Networks}, vol.~7, pp.~455--466, 2001.

\end{thebibliography}
\end{document}